\documentclass[%
 reprint,
 amsmath,amssymb,
 aps,
]{revtex4-1}

\usepackage{graphicx}
\usepackage{dcolumn}
\usepackage{bm}

\begin{document}

\preprint{APS/123-QED}

\title{SI based disease model over signed network }

\author{Cong Wan}
\email{10000cong@163.com}
\author{Cong Wang}
\author{Yanxia Lv}
\affiliation{%
School of Computer and Communication Engineering, NorthEastern University at Qinhuangdao, Hebei, China
}%

\date{\today}

\begin{abstract}
Signed network is a kind of network that associates each edge a positive or negative sign which could express friendly or unfriendly relationship between individuals. When diseases spreading over the signed network, some unfriendly edges may refuse to spread. Moreover, the signed network is dynamic, according to the structure balance theory which is the most important theory in the study of signed network, edges in the signed network will flip their signs over time. How does disease spreading interact with signed network evolving becomes a challenging issue. In this paper, we propose an energy function to describe the disease spreading and network evolving together, and we introduce the notion of Structure and Spreading Balanced. We extend the structure balance theory of Cartwright and propose a Structure and Spreading theorem. Finally, we carry out Monte-Carlo simulations on complete signed network to validate our theorem. In the experiment, we find that the signed network has self-immunity during disease spreading, which can be used explain the phenomenon of isolating the virulent virus by isolation.
\begin{description}

\item[PACS numbers]
87.23.Kg, 87.19.xw, 02.50.Ng
\end{description}
\end{abstract}

\pacs{Valid PACS appear here}
\maketitle


\section{instruction}

Recently years, research on disease spreading model has becoming an important research field. A number of spreading models have been proposed, such as SI\cite{4} and SIR\cite{Li2014An}. According to these models, disease or message spreading over complex networks from nodes to nodes through edges. However, sometimes nodes refuse to accept message through some edges\cite{4}, e.g., people refuse to receive messages form their enemies in social network. That means friendly or unfriendly relationship between nodes could affect disease spreading, which has not been deeply researched yet. Moreover, friend or enemy may exchange their roles\cite{Guo2017How}, which makes the process of disease spreading more complex.

Signed network is kind of network that associate each edge a sign (positive or negative) indicating whether two nodes are friends or enemies. A basic characterization of signed network is the notion of structure balance\cite{1}. Firstly, it was used to define whether a triple which consist of three interconnected nodes is balanced, e.g., two nodes hate each other, but the third node is their common friend, then their relationship may be Embarrassing, so this triple is unbalanced. Then Cartwright and Harary\cite{2} studied the balanced network which indicates all the triples in the network are balanced. They proposed structure balance theory and showed how to identify a balanced network at a glance, i.e., a structure balanced network can be divided into subgroups with positive links inside each subgroup and negative links among different subgroups. Structure balance theory has since become an important topic in the study of signed networks. Davis\cite{weakBalance}proposed a generalized structural balance theory which relax the conditions of balanced. It considered a triple with three mutually hostile nodes as balanced.
Moreover, many researchers have put forward further researches. Jurgen Lernery\cite{Lerner2016Structural} proposed a model to estimate the probability of friendly or unfriendly relationship. His model used more explanatory variables rather than only three variables in balanced theory.Haifeng Du et al. \cite{Du2016Structural}proposed a new form of structure balance which not only consider the sign of edges but also consider the sign of nodes.

A large signed network is always unbalanced if the relationships of its nodes are randomly chosen to be friendly or unfriendly\cite{3}. An unbalanced network is tending to flip the sign of its edges, which makes the network evolving from unbalanced state to balanced state, e.g., two people hate each other, but they have a common friend, then their relationship may be improved to make this triple stable. The empirical analysis of real network confirms balance theory, many works\cite{Leskovec2010Signed}\cite{Szell2010Multirelational} showed that the number of balanced triples is more than the number of unbalance ones.
Many methods have been proposed to describing the evolution of signed network, such as Energy landscape based models\cite{3}\cite{4}\cite{5}\cite{6} and information entropy based model\cite{Guo2017How}. Form the energy view, the more balanced triple the network has, the less energy this network gets. A balanced signed network has the lowest possible energy. Therefore, the evolution of signed networks is accompanied by a decline in energy. The process of evolution can be simulated using Monte Carlo methd\cite{4}\cite{7} or continuous methd\cite{8}\cite{9}.

Opinion Dynamic is also an important topic in the study of signed network. Individuals in the network could hold a binary (somebody supports something or not) or continues (the degree somebody supports or not supports something) opinion. When they communicate with each other, their opinion may be changed, and the friendly or unfriendly relationships between individuals would influence the opinion\cite{10}\cite{11}\cite{12}\cite{13}\cite{14}. Altafini\cite{10} assumed the network is structure balanced and studied how opinion evolves over such network, his work showed monotone dynamical systems could be used to describe how the opinions formed. Daron Acemoglu et al. presented models for binary or continuous opinion dynamic over a network with ¡°"stubborn agents¡± that never change their states\cite{11}\cite{12}, and they proved opinion could not converge in such network.  Shi et al. \cite{13}presented a DeGroot \cite{14} based non-Bayesian model and a lemma referred as ¡°live-or-die¡± lemma, his work showed how opinion evolves over signed network and proved the opinion will either converge to an agreement or diverge. Traag et al.\cite{Traag2013Dynamical} presented a model that update individuals' opinion based on the homogeneous process. According to their work, if an individual A want to revise his opinion about someone B, he will ask B's opinion about everyone else instead of asking everyone's opinion about B.

The process of signed network evolution has been widely studied, but the following problems remaining to be solved: (1) how does a signed network evolving together with disease spreading? (2) what impact dose signed network evolution have on disease spreading?
In this paper, we are trying to tackle such issues. Our work is based on SI model which is one of the most common spreading models. For SI model, each node in the network has two possible states which are susceptible and infected. A susceptible node may be influenced by its infected neighbor and switch to infected state, but an infected node will never be changed. However, message or disease cannot spread from an unfriend neighbor, e.g., you may not trust or contact with your enemy, so you will avoid being influenced\cite{4}\cite{Acemoglu2010Spread}. It means negative edges are unavailable for disease spreading. Given an unbalanced signed network, if some nodes are initialized to be infected, then how does disease spreading interactive with signed network evolution will be a problem. Our approach is to define a novel energy function which considers both structure energy and spreading energy. According to our energy function, the balanced network with the lowest energy are structure balanced and spreading balanced. It means all the triples in the network are balanced and disease cannot spread anymore. Furthermore, we propose structure and spreading theory which extend structure balance theory. It defines whether a signed network is structure and spreading balanced.

The rest of this paper is organized as follows: In Section 2, we introduce our energy function and structure and spreading theory. Section 3 presents the simulation method. Section 4 presents the results of simulations. Section 5 presents some concluding remarks.

\section{disease spreading and network evolution model}
\subsection{A Signed network model and structural balance theory}

\begin{figure}
\includegraphics[width=6cm]{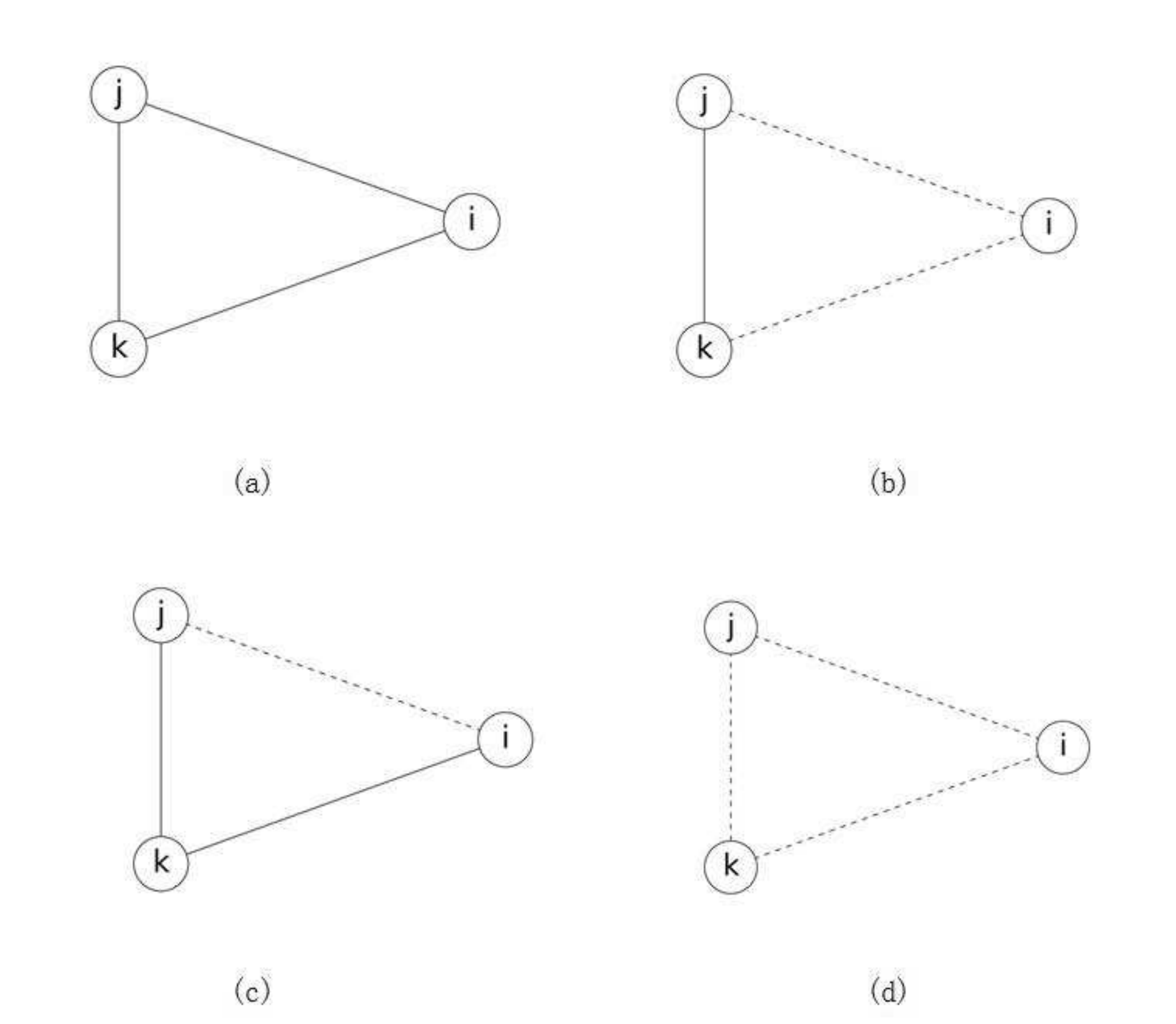}
\caption{\label{fig:1} Relationship patterns of triples in signed network.}
\end{figure}

A complete signed network can be modeled as a undirected weighted graph $G = (V,A)$. Here $V = \{ 1,2,3...n\}$ is a finite set of nodes, and \emph{A} is an adjacency matrix. If the edge between nodes \emph{i} and \emph{j} is positive, then $A_{ij}^{} = 1$. If the edge between nodes \emph{i} and \emph{j} is negative, then $A_{ij}^{} = -1$. A triple (\emph{i}, \emph{j}, \emph{k}) is formed by three distinct nodes in \emph{V}.
According to Heider's work\cite{2}, The balanced or unbalanced status of a triple (\emph{i}, \emph{j}, \emph{k}) is due to the structure energy contribution $H(i,j,k) =  - A_{ij}^{}A_{ik}^{}A_{ki}^{}$, as $H =  + 1( - 1)$ represents that the triple is unbalanced or balanced. A triple can have four possible configurations shown in Fig.~\ref{fig:1}., in which solid and dashed edges show positive and negative relationships respectively. Triples with configuration Fig.~\ref{fig:1} (a) and (b) are balanced, those with configuration Fig.~\ref{fig:1} (c) and (d) are unbalanced. If all the triples in \emph{G} is balanced, then the signed network \emph{G} is structural balanced. The total structure energy contribution of the network $E_e$ is Eq.~(\ref{eq:1}). When all the triples in the network is balanced, the structure energy is -1, thus the network is structure is balanced.
\begin{eqnarray}
 E_e^{} =  - \frac{1}{{(_3^n)}}\sum\limits_{ijk} {A_{ij}^{}A_{ik}^{}A_{ki}^{}}
 \label{eq:1}.
\end{eqnarray}

An unbalanced triple is tending to flip the sign of its edges, until it becomes balanced. However, in the network, an edge is always shared by multiple triples. The flipping of an edge could change the configuration of all these triples, which makes the total structure energy increase or decrease. So, whether or not to accept the flipping depends on whether total structure energy is reduced.

\subsection{SI model and spreading energy}
\begin{figure}
\includegraphics[width=6cm]{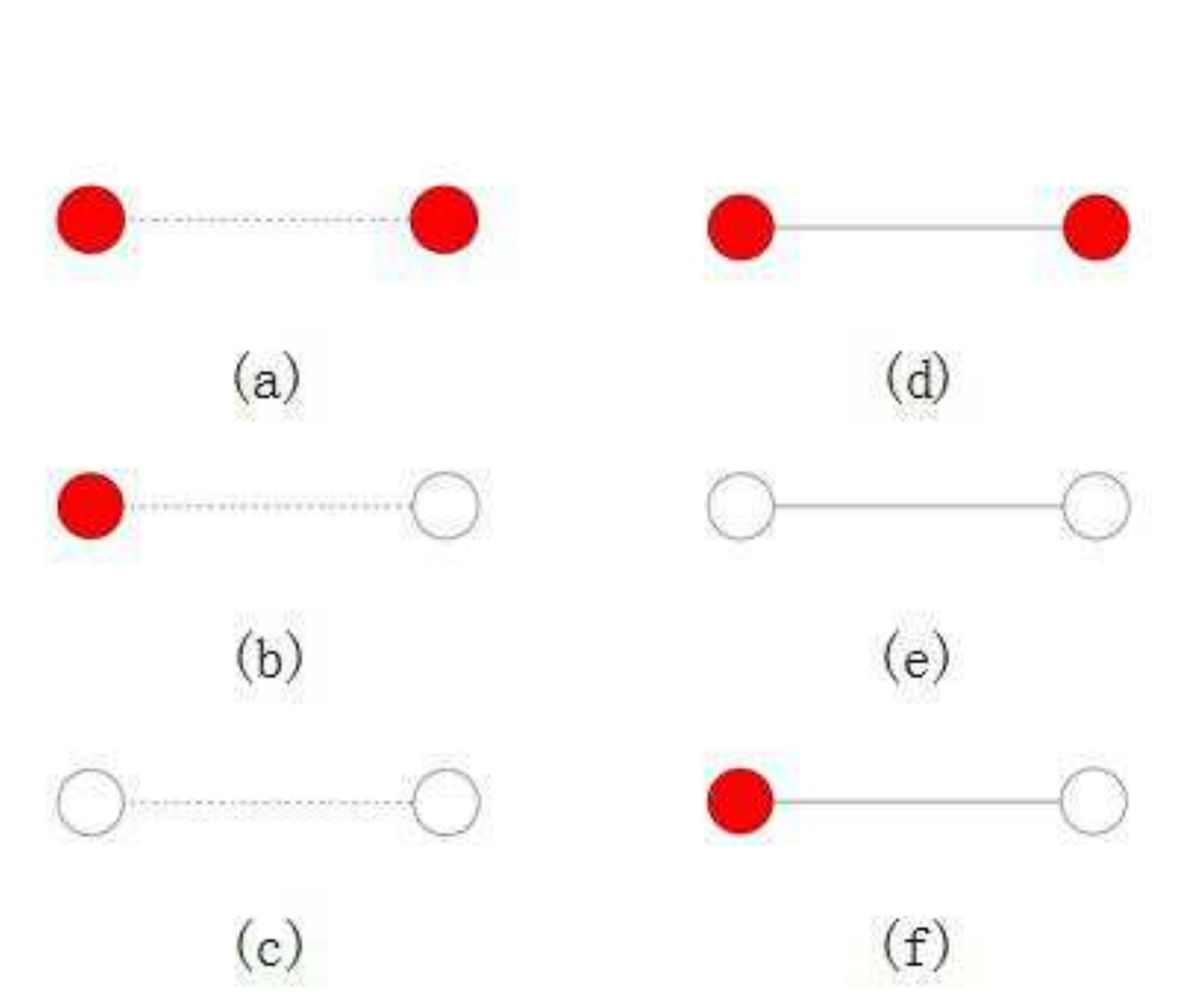}
\caption{\label{fig:2} Six different states of edges and its end nodes.}
\end{figure}

The SI model is used to describe diseases that are incurable, we assume diseases spread on the signed network following the SI model. Let $S_i^{}$ be the state of node \emph{i} in the signed network. If the node is susceptible, then $S_i^{} = 1$. If the node is infected, then $S_i^{} = -1$. Let \emph{M} be the number of susceptible nodes in the network, and let \emph{N} be the number of infected nodes in the network. Fig.~\ref{fig:2}. lists Six different configurations of edge and its end nodes, in which red and white nodes show infected and susceptible state respectively.

Considering the disease spreading may affect the evolution of signed network, we defined an energy function for each configurations of edge. We referred this energy as spreading energy. Spreading energy of the edge between nodes \emph{i} and \emph{j} is represented by $E_{ij}$ which described as follow:

\begin{eqnarray}
E_{ij}^{} = A_{ij}^{}\frac{{(S_i^{} - S_j^{})_{}^2}}{4}
 \label{eq:2}.
\end{eqnarray}

We assume diseases spreading only through the positive edges. The configuration of edges and endpoints can be divided into three categories:

(1) No matter the edge is positive or negative, there is no disease spreading behavior. Such as (a)(c)(d)(e) in Fig.~\ref{fig:2}, the disease doesn¡¯t spread even if the sign of edge flips. These configurations have no contribution to network balance or unbalance, so their spreading energy is 0.

(2) The disease can't spread, but if the sign of edge flips, disease may spread, such as (b) in Fig.~\ref{fig:2}. The spreading energy of this configuration is -1. Such edge is referred as balanced edge.

(3) The disease may spread, such as (f) in Fig.~\ref{fig:2}. The spreading energy of this configuration is 1. Such edge is referred as unbalanced edge.

The total spreading energy $En$ of the signed network described as Eq.~(\ref{eq:3}). If the network is spreading balanced, i.e., the disease can't spread in the network anymore, then $En = -1$.

\begin{figure}
\includegraphics[width=8cm]{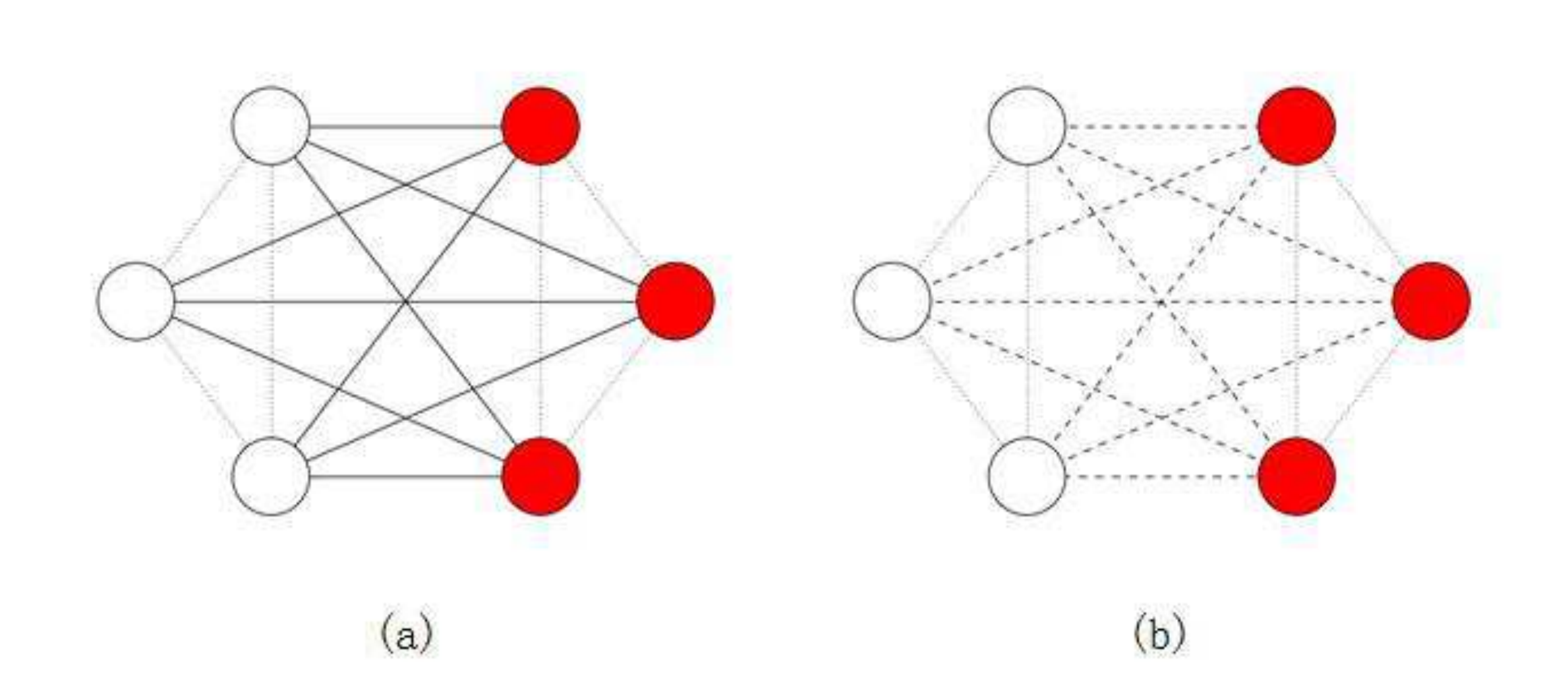}
\caption{\label{fig:3} Illustrations of the signed network with highest spreading energy and lowest spreading energy.}
\end{figure}

Fig.~\ref{fig:3}(a) and (b) show illustrations of signed network with maximum and minimum spreading energy. In Fig.~\ref{fig:3}, red and white nodes show infected and susceptible nodes, solid and dashed lines show positive and negative relationships respectively, and the dotted line indicates the relationship can be both negative and positive. Nodes in these networks can be divided into two subsets. One subset only contains infected nodes, meanwhile, the other one only contains susceptible nodes. In Fig.~\ref{fig:3} (a), there are only positive edge between two subsets, wherever, in Fig.~\ref{fig:3}(b), there are only negative edge between two subsets.

\begin{eqnarray}
En = \left\{ \begin{array}{l}
\frac{1}{2}*\frac{{\sum\limits_{ij} {E_{ij}^{}} }}{{M*N}}{\kern 1pt} {\kern 1pt} {\kern 1pt} {\kern 1pt} {\kern 1pt} {\kern 1pt} {\kern 1pt} {\kern 1pt} {\kern 1pt} ,if{\kern 1pt} {\kern 1pt} {\kern 1pt} {\kern 1pt} M*N \ne 0\\
 - 1{\kern 1pt} {\kern 1pt} {\kern 1pt} {\kern 1pt} {\kern 1pt} {\kern 1pt} {\kern 1pt} {\kern 1pt} {\kern 1pt} {\kern 1pt} {\kern 1pt} {\kern 1pt} {\kern 1pt} {\kern 1pt} {\kern 1pt} {\kern 1pt} {\kern 1pt} {\kern 1pt} {\kern 1pt} {\kern 1pt} {\kern 1pt} {\kern 1pt} {\kern 1pt} {\kern 1pt} {\kern 1pt} {\kern 1pt} {\kern 1pt} {\kern 1pt} {\kern 1pt} {\kern 1pt} {\kern 1pt} {\kern 1pt} {\kern 1pt} {\kern 1pt} {\kern 1pt} {\kern 1pt} {\kern 1pt} {\kern 1pt} {\kern 1pt} {\kern 1pt} {\kern 1pt} {\kern 1pt} {\kern 1pt} ,if{\kern 1pt} {\kern 1pt} {\kern 1pt} {\kern 1pt} M*N = 0
\end{array} \right.
 \label{eq:3}.
\end{eqnarray}

\subsection{Structure and Spreading theorem}
We define the total energy \emph{E} of the signed network as the weighted sum of structure energy and spreading energy, as Eq.~(\ref{eq:4})

\begin{eqnarray}
E = \alpha *Ee + (1 - \alpha )*En
 \label{eq:4}.
\end{eqnarray}

If the network is fully balanced, then $E =  - 1$, which indicates that all the triples and edges are balanced, while $E = 1$ indicates that all the triples and edges of the network are unbalanced. In order to determine whether or not a given network is balanced without check every triples and edges, we propose the Structure and Spreading theorem.

\textbf{Structure and Spreading theorem:}

If a signed network $G(V,A)$ satisfies the following conditions, then \emph{G} is balanced.

(1) The nodes of \emph{G} can be separated into two clusters \emph{X} and \emph{Y}, in which $X \cup Y = V$ and $X \cap Y = \emptyset$

(2) For $\forall i,j \in X$ or $\forall i,j \in Y$, the $A_{ij}^{} = 1$. It indicates that all the edges in one cluster are positive.

(3) For $\forall i \in X$ and $\forall j \in Y$, the $A_{ij}^{} = -1$. It indicates that all the edges between two clusters are negative.

(4) For $\forall i \in X$, the $s_i^{} = 1$. It indicates that all the nodes in cluster \emph{X} are susceptible.

(5) For $\forall j \in Y$, the $s_i^{} = -1$. It indicates that all the nodes in cluster \emph{Y} are infected.

\begin{figure}
\includegraphics[width=8cm]{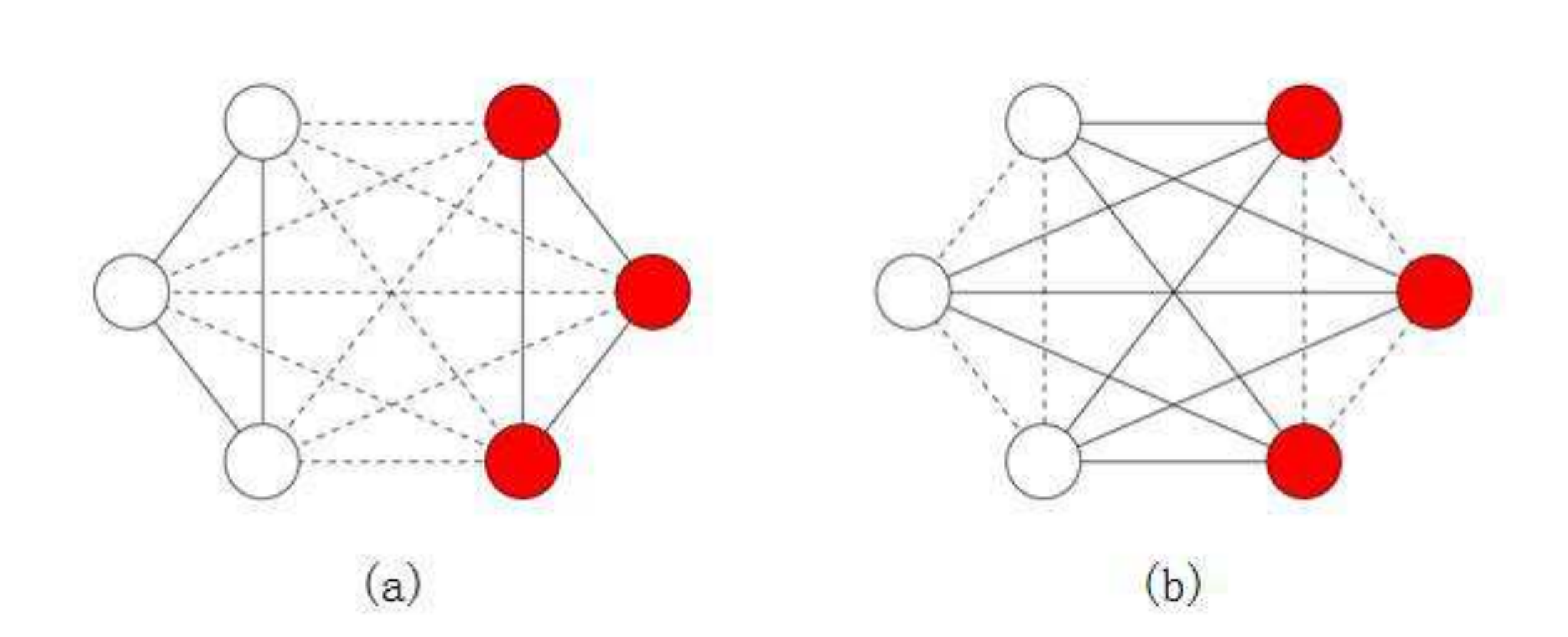}
\caption{\label{fig:4} Illustrations of the balanced signed network and the most unbalanced signed network according to spreading and structure theorem.}
\end{figure}

As shown in Fig.~\ref{fig:4}(a), the nodes are separated into two clusters that only contains infected or susceptible nodes respectively such that each positive edge joins two nodes of the same cluster and each negative edge joins two nodes of the different cluster. Using Structure and Spreading theorem, we can immediately see that the network in Fig.~\ref{fig:4}(a) is balanced. In other way, according to Eq.~(\ref{eq:4}), the total energy of Fig.~\ref{fig:4}(a) is -1, so it's balanced. In contrast, the nodes in Fig.~\ref{fig:4}(b) are separated into two clusters that only contains infected or susceptible nodes respectively such that each negative edge joins two nodes of the same cluster and each positive edge joins two nodes of the different cluster. According to Eq.~(\ref{eq:4}), the total energy of Fig.~\ref{fig:4}(a) is -1, each triple and edge in this network is unbalanced.

\section{simulation}

Using Monte Carlo method, we can simulate the evolution of a signed network. There can be two operations in each step of the evolution, one is flipping the sign of a randomly selected edge, the other is trying to spread disease through this edge. Specifically, the evolution follows the following four rules:

(1) If end nodes are both susceptible or infected, then the sign of the edge flips.

(2) If the states of two end nodes are different and the edge is negative, then the sign of the edge flips.

(3) If the states of two end nodes are different and the edge is positive, then the disease spreads with probability $\lambda$.

(4) If the states of two end nodes are different and the edge is positive, then the disease doesn't spread with probability $1-\lambda$ and the sign of the edge flips.

In the process of Monte Carlo simulation, an edge is randomly selected, then the network \emph{G} evolves according to the above rules and we get a new network \emph{G'}. If the energy of \emph{G'} is less than the energy of \emph{G}, then this step of evolution is accepted and set \emph{G=G'}. Otherwise, this step of evolution will be cancelled. We start the evolution of the network from an initial state. Each node initialized to infected with probability $\rho$ and Initialized to susceptible with probability $1-\rho$. Each edge is initialized to positive with probability $\sigma$ and Initialized to negative with probability $1-\sigma$. The evolution process continues until the network reaches either following conditions. (1) The network is balanced, i.e., the energy is -1. (2) The network reaches a jammed state which means if we change state of any edge or node, the energy of network will be increased.

\section{Experiment}
\begin{figure*}
\includegraphics[width=18cm]{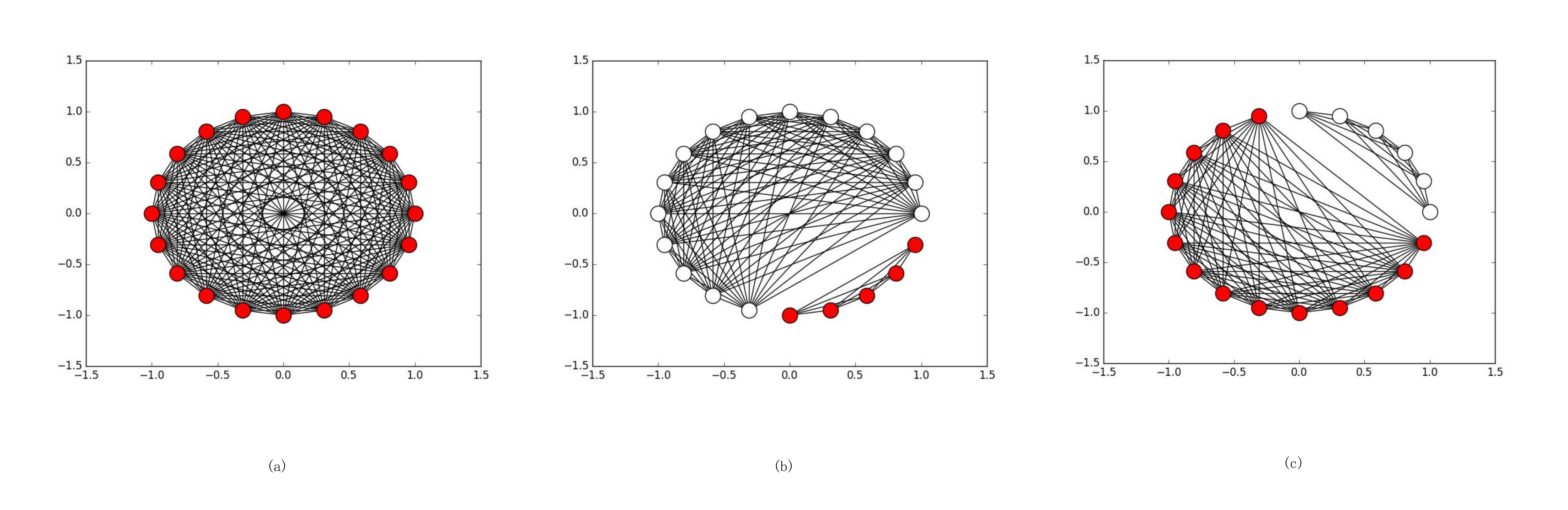}
\caption{\label{fig:5} Balanced signed network with 20 nodes after evolving. (a)$\sigma=0$, $\rho=0.6$, $\lambda=0.8$; (b)$\sigma=0$, $\rho=0.2$, $\lambda=0.2$; (c)$\sigma=0$, $\rho=0.4$, $\lambda=0.8$}
\end{figure*}

We initialize some signed networks with 20 nodes, then start the evolution using Monte Carlo method.  Fig.~\ref{fig:5} shows the jammed state and initialization parameters of such signed networks. In Fig.~\ref{fig:5}, red and white nodes show infected and susceptible node, and solid lines show positive relationships. In order to show the clusters clearly, the negative edges are not shown. Those nodes are divided into two clusters: one is made up of infected nodes, the other one is made up of susceptible nodes. Fig.~\ref{fig:5}(a) can also be considered as two clusters, in which susceptible cluster is empty. It can be observed that the spreading of disease is stopped by the evolution of signed network.

These illustrations are initialized with different spreading probability $\lambda$, initial positive edges ratio $\sigma$ and initial infected nodes ratio $\rho$. According to their initialization parameters, Fig.~\ref{fig:5}(a) has 12 infected nodes before evolution, and has 20 infected nodes after evolution; Fig.~\ref{fig:5}(b) has 4 infected nodes before evolution, and has 5 infected nodes after evolution; Fig.~\ref{fig:5}(c) has 8 infected nodes before evolution, and has 14 infected nodes after evolution. There are 8 nodes, 1 nodes and 6 nodes are infected respectively in such three configurations respectively. It indicates that initialization parameters could affect disease spreading result.

\begin{figure}
\includegraphics[width=8cm]{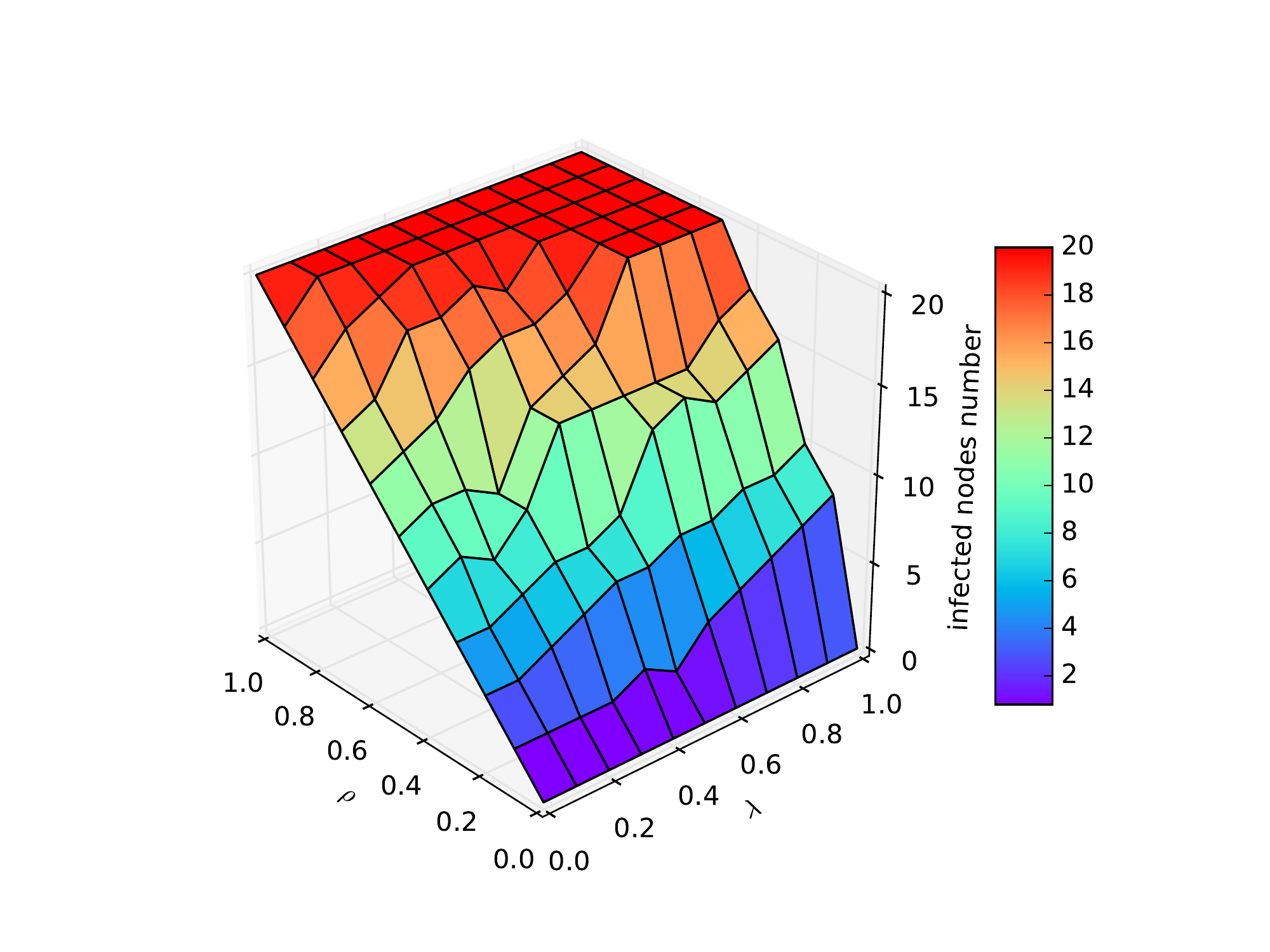}
\caption{\label{fig:6} Size of infected cluster with different $\rho$ and $\lambda$, in which $|V| = 20$, $\alpha  = 0.5$ and $\sigma  = 0.5$.}
\end{figure}

\begin{figure}
\includegraphics[width=8cm]{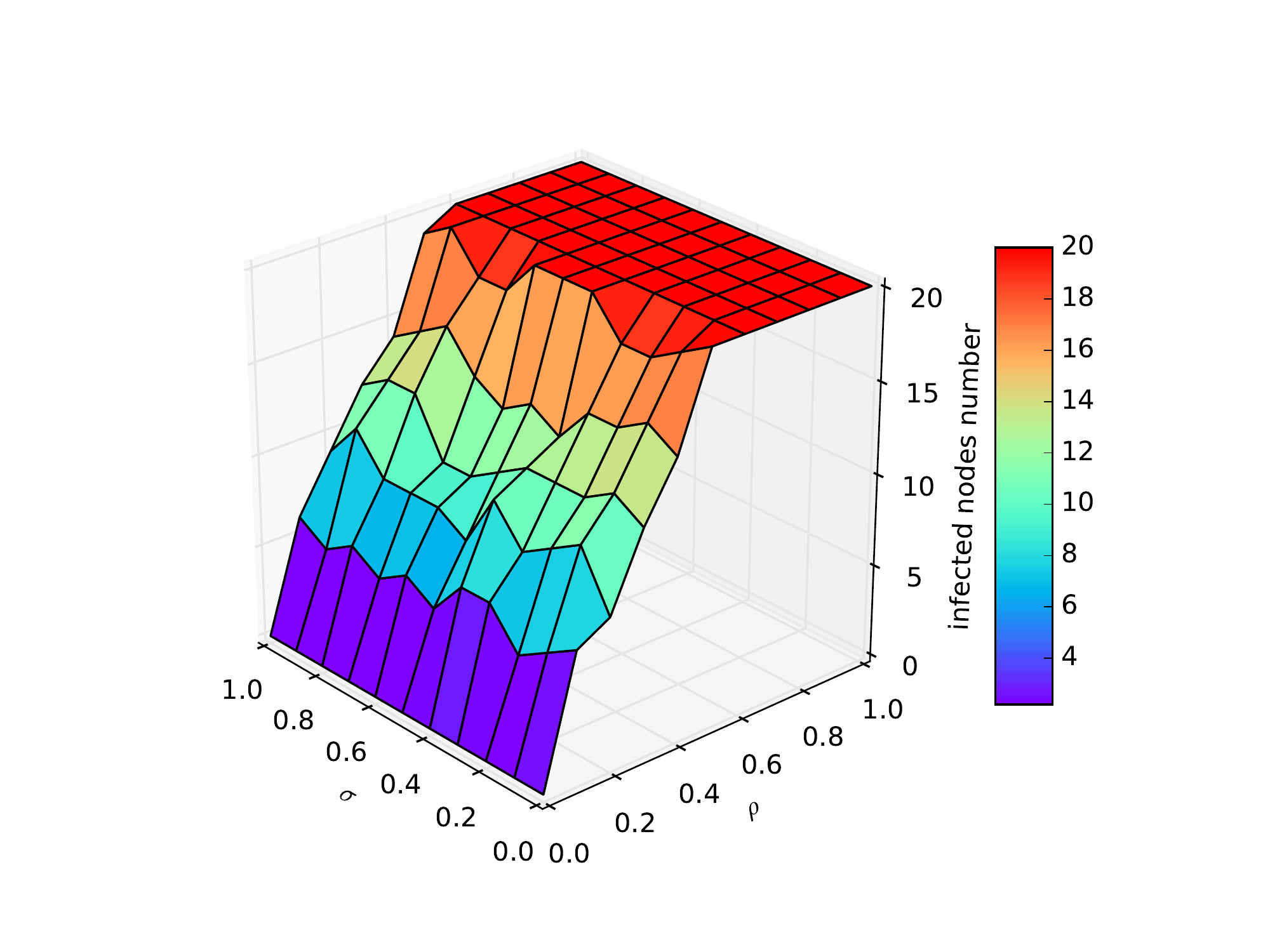}
\caption{\label{fig:7} Size of infected cluster with different $\sigma$ and $\rho$, in which $|V| = 20$, $\alpha  = 0.5$ and $\lambda  = 0.8$.}
\end{figure}

\begin{figure}
\includegraphics[width=8cm]{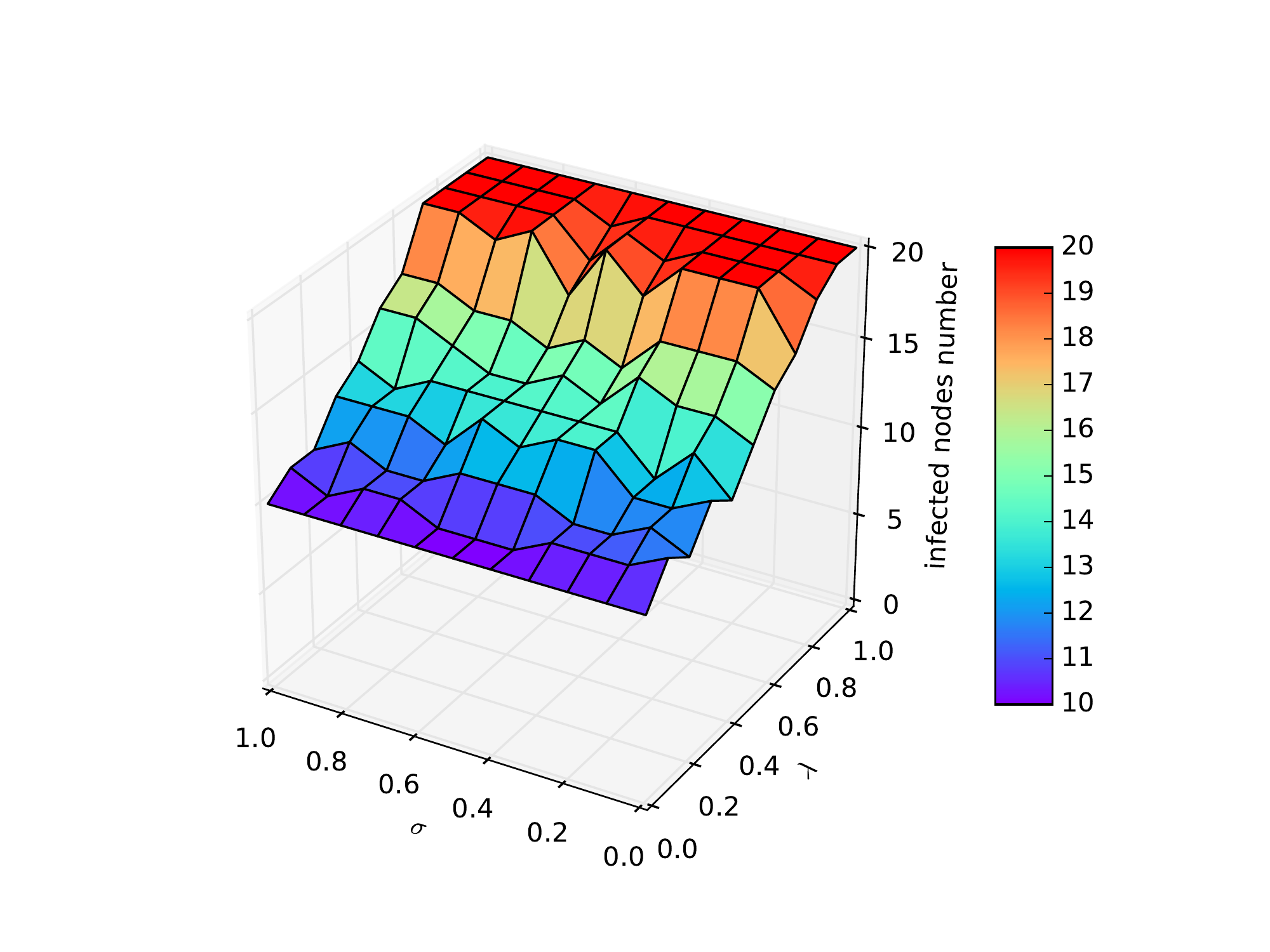}
\caption{\label{fig:8} Size of infected cluster with different $\sigma$ and $\lambda$, in which $|V| = 20$, $\alpha  = 0.5$ and $\rho  = 0.5$.}
\end{figure}

Fig.~\ref{fig:6} shows the number of infected nodes at balanced or jammed state with different $\rho$ and $\lambda$. It can be observed that the number of infected nodes increases with $\rho$ and $\lambda$, which indicates more initial infected nodes or higher spreading probability makes the infected cluster bigger. In another word, if the values of $\rho$ and $\lambda$ are increased, the process of disease spreading will be speed up. If the disease spreads too fast, then all the nodes will be infected as Fig.~\ref{fig:6}(a). Fig.~\ref{fig:7} shows the number of infected nodes at balanced or jammed state with different $\sigma$ and $\rho$. Fig.~\ref{fig:8} shows the number of infected nodes at balanced or jammed state with different $\sigma$ and $\lambda$. It can be observed that $\sigma$ has little effect on infected nodes number. Even all the edges are initialized to be negative, the disease will be still spread.

\begin{figure}
\includegraphics[width=8cm]{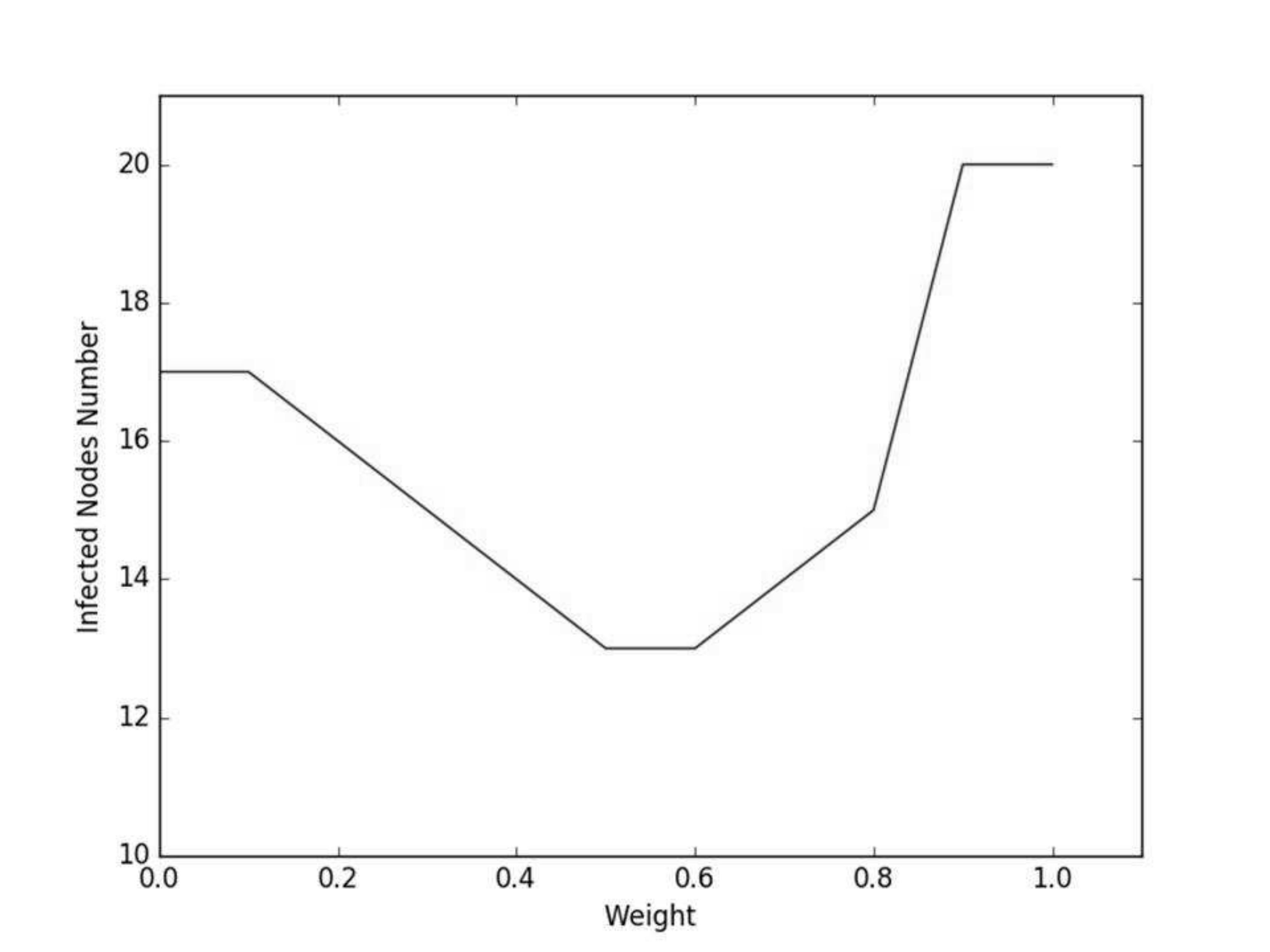}
\caption{\label{fig:9} Infected nodes number with different energy weight.}
\end{figure}

According to the definition of total energy Eq.~(\ref{eq:4}), with the increase of $\alpha$, the structure energy becomes more important and the spreading energy becomes less important.  Fig.~\ref{fig:9} shows the infected nodes number with different weights $\alpha$, in which all the results are averages of ten time execution. It indicates that increasing the importance of either the structure energy or the spreading energy can cause more infections.

\section{Conclusion}
In order to describe the interaction of signed network evolution and disease spreading process over the signed network. We have proposed structure and spreading theorem which extends Cartwright's\cite{2} theorem. We have introduced structure energy and spreading energy of the signed network and combined them as the total energy of the network. In this case, the network with total energy -1 is considered as balanced. According to our Structure and Spreading theorem, we can identify a balanced network without massive calculation. That is, if a signed network can be divided into two clusters which only contain susceptible or infected nodes, and edges between clusters are all negative and edges in one clusters are all positive, then the network is balanced. We initialize signed networks with parameters $\rho$ and $\lambda$ which represent the initial ratio of infected nodes and positive edges respectively. Then, we have started evolution using Monte Carlo method and showed the balanced networks which are divided into two clusters. Disease spreads over the network until all the infected nodes are separated from susceptible nodes by negative edges. The simulation have showed that higher $\rho$ and $\lambda$ makes more nodes infected. Furthermore, our experiments also shows that the evolution process makes the network immune to disease, i.e., a set of susceptible nodes are separated from the infected ones. It just like the segregation of infectious patients in the real society, e.g., the segregation to avoid SARS.

In the further, we will extend our work to SIS and SIR model. Furthermore, we will test our model on the real signed network, such as Epinions and Slashdot.
\begin{acknowledgments}
Acknowledges support from the National Natural Science Foundation of China (Grant No. 61501102, 61702089), and the Natural Science Foundation of Hebei,China(Grant No. F2016501079).
\end{acknowledgments}

\bibliography{apssamp}

\end{document}